\documentclass[noshowpacs,twocolumn,aps,hyperref,floatfix,raggedbottom,final,natbib]{revtex4-1}
\usepackage{amssymb,amsfonts,amsmath}
\usepackage{natbib}
\usepackage{euscript}
\usepackage{graphics}
\usepackage{setspace}
\usepackage{graphicx}
\usepackage{setspace}
\usepackage{dcolumn}
\usepackage{bm}
\usepackage{mathrsfs}
\usepackage{units}
\usepackage[pdfpagemode=UseNone]{hyperref}
\usepackage[usenames,dvipsnames]{color}
\usepackage[normalem]{ulem}
\usepackage{soul}
\usepackage{cancel}

\newcounter{defcounter}
\newenvironment{myequation}{%
\addtocounter{equation}{-1}
\refstepcounter{defcounter}
\renewcommand\theequation{{\color{blue}A\thedefcounter}}
\begin{equation}}
{\end{equation}}

\newenvironment{mysubequations}{%
\addtocounter{equation}{-1}
\refstepcounter{defcounter}
\renewcommand\theequation{{\color{blue}A\thedefcounter}}
\begin{subequations}}
{\end{subequations}}

\newcommand{\stkout}[1]{\ifmmode\text{\sout{\ensuremath{#1}}}\else\sout{#1}\fi}

\renewcommand\theequation{{\color{blue}\arabic{equation}}}

\hypersetup{colorlinks=true, citecolor=blue, linkcolor=blue, urlcolor=blue}

\newcommand\numberthis{\addtocounter{equation}{0}\tag{\theequation}}

\allowdisplaybreaks

\begin{document}

\title{Alternative mechanism for coffee-ring deposition based on active role of free surface}

\author{Saeed Jafari Kang,$^1$ Vahid Vandadi,$^1$ James D. Felske,$^2$ and Hassan Masoud$^{1,}$}
\email[Email address: ]{hmasoud@unr.edu}

\affiliation{$^1$Department of Mechanical Engineering, University of Nevada, Reno, Nevada 89557, USA\\
$^2$Department of Mechanical and Aerospace Engineering, State University of New York, Buffalo, New York 14260, USA}

\date{\today }

\begin{abstract}
When a colloidal sessile droplet dries on a substrate, the particles suspended in it usually deposit in a ring-like pattern. This phenomenon is commonly referred to as the ``coffee-ring'' effect. One paradigm for why this occurs is as a consequence of the solutes being transported towards the pinned contact line by the flow inside the drop, which is induced by surface evaporation. From this perspective, the role of the liquid-gas interface in shaping the deposition pattern is somewhat minimized. Here, we propose an alternative mechanism for the coffee-ring deposition. It is based on the bulk flow within the drop transporting particles to the interface where they are captured by the receding free surface and subsequently transported along the interface until they are deposited near the contact line. That the interface captures the solutes as the evaporation proceeds is supported by a Lagrangian tracing of particles advected by the flow field within the droplet. We model the interfacial adsorption and transport of particles as a one-dimensional advection-generation process in toroidal coordinates and show that the theory reproduces ring-shaped depositions. Using this model, deposition patterns on both hydrophilic and hydrophobic surfaces are examined in which the evaporation is modeled as being either diffusive or uniform over the surface.
\end{abstract}

\maketitle

\section{Introduction}

Understanding why a spilled drop of coffee leaves behind a ringlike stain after it dries out \cite{Deegan1997} might have been a matter of academic curiosity at first. However, soon, it was found that the lessons learned from studying the so-called ``coffee-ring'' effect and phenomena related to it have important implications in ink-jet printing ordered structures via the evaporative self-assembly technique (see Ref. \cite{Han2012} and references therein). This has motivated many researchers to investigate mechanisms underlying the deposition of nonvolatile solutes in evaporating sessile drops \cite{Deegan2000-a,
Deegan2000-b,
Fischer2002,
shmuylovich2002,
Pauchard2003-a,
Popov2005,
Heim2005,
lin2005,
Hu2006,
bigioni2006,
Smalyukh2006,
Yarin2006,
schnall2006,
Widjaja2008,
Bhardwaj2009,
bhardwaj2010,
joshi2010,
yunker2011,
marin2011,
still2012,
hampton2012,
larson2012,
das2012-a,
das2012-b,
majumder2012,
yunker2013,
sempels2013,
nguyen2013,
chen2013,
Breinlinger2014,
anyfantakis2014,
wray2014,
Devlin2015-b,
anyfantakis2015,
boulogne2015,
erbil2015,
du2015,
pack2015,
Tarasevich2016,
Li2016,
karapetsas2016,
kim2016}. A review of recent studies can be found in Ref.~\cite{Larson2014}.

The majority of investigations to date have assumed (or concluded) that the particle deposition is a consequence of the evaporation-induced bulk flow transporting particles either directly to the substrate or to a pinned contact line. By contrast, fewer studies have considered that the bulk flow carries particles to an inevitable intersection with the liquid-gas interface (see,  e.g., Refs.~\cite{Li2016,anyfantakis2015,Breinlinger2014,anyfantakis2014,yunker2013,yunker2011,bigioni2006,Deegan2000-a,Deegan2000-b}). Here, we demonstrate that this inevitable intersection results from the coupling of the receding of the interface with the bulk internal flow. We also reexamine the particle deposition during the drying of a colloidal sessile droplet (e.g.,~a drop of coffee) by explicitly taking into account the intersection of nonvolatile suspended particles with the liquid-gas interface.

We propose that upon arriving at the interface the particles move with the velocity of the surface, wherein the tangential component of the fluid flow carries them towards the contact line. This assumption splits the transport of solutes into interfacial and bulk. We model the interfacial adsorption and transport of particles as a one-dimensional advection-generation process in toroidal coordinates. We show, perhaps surprisingly, that, if the solutes are initially distributed uniformly within the drop (the usual case), then their bulk concentration remains uniform during evaporation. Physically, their capture by the interface is counterbalanced by the reduction in drop volume due to evaporation. Therefore, the consideration of interfacial transport alone is sufficient for determining where on the substrate the particles will finally reside.

We numerically solve the transport equation for the evolution of particle concentration at the interface. Indeed, we find that the results predict ringlike deposition patterns and they quantitatively agree with available experimental data. These findings raise the possibility of an alternative mechanism for the coffee-ring effect in which the free surface plays an active role in the particle deposition. In what follows, we will first describe our mathematical model and its numerical solution and then present the results for drops of various contact angles subject to diffusive and uniform evaporation profiles.

\begin{figure*}
 \includegraphics[width = \textwidth]{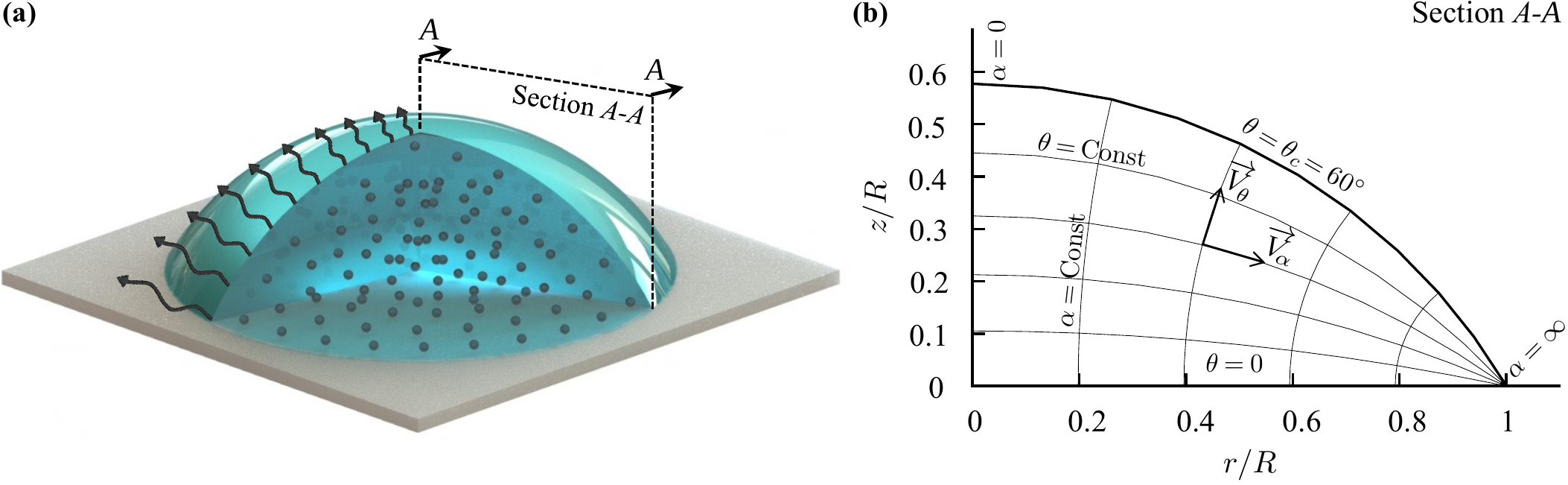}
 \caption{(Colour online) (a) An evaporating sessile droplet containing a dilute suspension of nonvolatile particles. (b) The lines of constant $\alpha$ and $\theta$ and positive direction of velocity components in a (left-handed) toroidal coordinate system that fits the boundaries of the drop on the meridian plane $A$-$A$.}
\label{figure1}
\end{figure*}

\section{Description of model}

Consider a colloidal droplet evaporating on a flat substrate [see Fig.~\ref{figure1}(a)]. To reduce the complexity of the analyses, the following (justifiable) assumptions are made:  (i) the droplet's shape is that of a spherical cap, (ii) the initial suspension of colloidal particles is uniform and dilute, (iii) the contact line remains pinned during evaporation until the contact angle has become very small, (iv) the particle diffusion is negligible compared to advection, (v) the particles are transported at the same velocity as the local solvent velocity, (vi) unlike the solvent, the particles do not evaporate (i.e.,~they remain in the solution), (vii) the volume occupied by a particle is negligible, (viii) the particles do not interact with one another nor interfere with the geometry of the drop or the flow field within it, (ix) the system is isothermal, and (x) Marangoni, gravitational, and inertial effects are negligible.

A typical experimental setup for which the assumptions are valid is a colloidal water droplet of initial diameter $D \sim 1\, \unit{mm}$ that evaporates under room conditions. Such a droplet becomes unpinned at $\theta_c = 2^{\circ}-4^{\circ}$ \cite{Hu2002}, and the characteristic flow velocity in its interior is $U = ( D_v / D )( \varphi_s - \varphi_\infty ) / \rho \sim 1 \, \mu \unit{m} / \unit{s}$ \cite{Hu2005-a}, where $D_v$ is the vapor diffusion coefficient, $\varphi_s$ is the vapor density at the droplet surface (saturation value), and $\varphi_\infty$ is the far field vapor density [see also Eq.~(\ref{eq:J_diff}) of the Appendix].

The capillary, Bond, and Reynolds numbers of this system are, respectively, $\mathrm{Ca} = \mu U / \gamma_{{}_0} \sim 10^{-8}$, $\mathrm{Bo} = \rho g D^2 / 4 \gamma_{{}_0} \sim 10^{-2}$, and $\mathrm{Re} = \rho D U / \mu \sim 10^{-3}$. Also, the P\'{e}clet number corresponding to the suspended particles of diameter $D_p \sim 1\, \mu\unit{m}$ and diffusion coefficient $\mathscr{D} \sim 10^{-13} \, \unit{m}^{2} / \unit{s}$ is $\mathrm{Pe} = D_p U / \mathscr{D} \sim 10^{2}$. The nondimensional numbers are calculated based on the water properties at room conditions (i.e., density of $\rho = 997.0479 \, \unit{kg} / \unit{m}^3$, viscosity of $\mu = 8.90 \times 10^{-4} \, \unit{N}\,\unit{s} / \unit{m}^2$, and surface tension $\gamma_{{}_0} = 71.97 \times 10^{-3} \, \unit{N} / \unit{m}$), and gravitational acceleration of $g = 9.81 \, \unit{m} / \unit{s}^2$. In this setup, the surface tension is strong enough to maintain the spherical cap shape during the evaporation ($\mathrm{Ca} \ll 1$ and $\mathrm{Bo} \ll 1$), the flow inside the droplet is creeping ($\mathrm{Re} \ll 1$), and the transport of the solutes is dominated by advection ($\mathrm{Pe} \gg 1$). Furthermore, thermal Marangoni effects are counterbalanced by a minute amount of surfactant contamination commonly present on water surfaces \cite{Hu2005-b}.

As the droplet dries out, three possibilities exist: (i) the particles arrive at the substrate, (ii) they intersect the liquid-gas interface, or (iii) they remain in and are transported by the bulk flow. The fluid velocity tends to zero as the substrate is approached. Hence, the suspended particles will not intersect the substrate without any significant attractive forces. On the other hand, the intersection with the liquid-gas interface is expected as a consequence of the velocity of the fluid always carrying the particles towards the free surface (see, e.g., Refs.~\cite{Masoud2009-a,Masoud2009-b}). To develop a proper mathematical model for the deposition, it is then crucial to know what fraction of particles intersect the interface during the life time of the drop. To this end, we carry out a series of numerical simulations where we track the motion of pointlike particles uniformly seeded inside evaporating droplets of various initial contact angles $\theta_{c_0}$ [see also Fig.~\ref{figure1}(a)]. We consider an axially symmetric evaporative flux resulting from a purely diffusive vapor phase transport into an infinite ambient [see Eq.~(\ref{eq:J_diff}) of the Appendix].

In the absence of diffusion, the suspended particles follow the fluid flow. Therefore, their trajectories can be computed by integrating
\begin{equation}\label{eq:disp}
\frac{d \boldsymbol{R}_i}{dt} = \boldsymbol{V} \left( \boldsymbol{R}_i \right),
\end{equation}
where $\boldsymbol{R}_i$ and $\boldsymbol{V} \left( \boldsymbol{R}_i \right)$ denote the position of the $i$th particle and the fluid velocity inside the drop at $\boldsymbol{R}_i$, respectively. We use the velocity field derived by Masoud and Felske for axisymmetric Stokes flow in evaporating sessile drops \cite{Masoud2009-b}. Briefly presented in the Appendix, their analytical solution is obtained in a toroidal coordinate system ($\alpha$,$\theta$,$\phi$), which exactly fits the boundaries of the drop. Here, we adopt the same coordinate system. Figure \ref{figure1}(b) shows a cross section of the droplet at a given azimuthal angle $\phi$ where the toroidal coordinates ($\alpha$,$\theta$) are indicated along with the cylindrical coordinates ($r$,$z$). The metric coefficients for the toroidal geometry are
\begin{equation}
h_{\alpha} = h_{\theta} = \frac{h_{\phi}} {\sinh{\alpha}} = \frac{R}{\cosh{\alpha} + \cos{\theta}},
\end{equation}
where $0 \le \alpha < \infty$, $0 \le \theta < 2 \pi$, $0 \le \phi < 2\pi$, and $R$ is the distance from the $z$ axis to the contact line. The relationships between the toroidal and the cylindrical coordinates are
\begin{equation}\label{coordinate_relations}
\frac{r}{\sinh{\alpha}} = \frac{z}{\sin{\theta}} = \frac{R}{\cosh{\alpha} + \cos{\theta}}.
\end{equation}

\begin{figure}
 \includegraphics[width = 0.485 \textwidth]{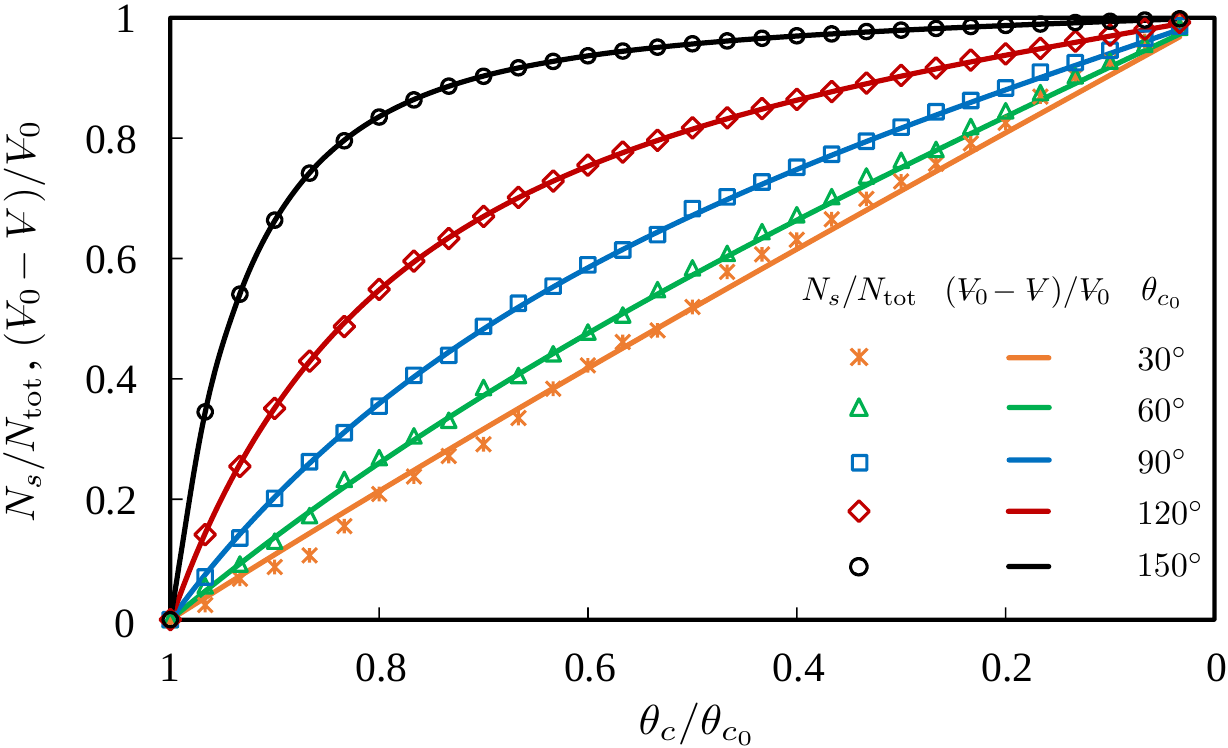}
 \caption{(Colour online) The cumulative fraction of particles captured by the liquid-gas interface $N_s / N_{\text{tot}}$ and the fraction of droplet's lost volume $\left(\stkout{V}_0 - \stkout{V} \right) / \stkout{V}_0$ versus $\theta_c / \theta_{c_0}$ for droplets of different initial contact angles. The symbols and solid lines represent $N_s / N_{\text{tot}}$ and $\left( \stkout{V}_0 - \stkout{V} \right) / \stkout{V}_0$, respectively.}
\label{figure2}
\end{figure}

Rewriting Eq. (\ref{eq:disp}) in the toroidal coordinates, we have
\begin{subequations}\label{disp_toroid}
\begin{align}
\frac{d \alpha_i}{d \theta_c} &= \frac {V_{\alpha}}{h_{\alpha} \dot{\theta}_c},\\
\frac{d \theta_i}{d \theta_c} &= \frac {V_{\theta}}{h_{\theta} \dot{\theta}_c},
\end{align}
\end{subequations}
where $\dot{\theta}_c = d \theta_c / dt$ is the rate of change of the contact angle, which itself is a function of $\theta_c$ (see also the Appendix). These equations are naturally nondimensional and can be integrated numerically using a second-order Adams-Bashforth method as
\begin{subequations}\label{disp_toroid_disc}
\begin{align}
\alpha^{n+1}_i &= \alpha^n_i + \frac{\Delta \theta_c}{2} \left[ 3 \left( \frac{h_{\alpha} V_{\alpha}}{\dot{\theta}_c} \right)^n_i -  \left( \frac{h_{\alpha} V_{\alpha}}{\dot{\theta}_c} \right)^{n-1}_i \right],\\
\theta^{n+1}_i &= \theta^n_i + \frac{\Delta \theta_c}{2} \left[ 3 \left( \frac{h_{\theta} V_{\theta}}{\dot{\theta}_c} \right)^n_i -  \left( \frac{h_{\theta} V_{\theta}}{\dot{\theta}_c} \right)^{n-1}_i \right],
\end{align}
\end{subequations}
where $\Delta \theta_c = \theta^{n+1}_c - \theta^n_c$ with the superscript $n$ being an integer. In our calculations, we set $\Delta \theta_c = -1/2^{\circ}$ and track the trajectory of the seeded particles until the contact angle is reduced to $\theta_c = 3^{\circ}$. We identify the particles that intersect the interface by checking whether $\theta^n_i \ge \theta^n_c$. At this stage, we no longer update the position of the particles that satisfy this condition. Lastly, the number of seeded particles $N_{\text{tot}}$ are chosen such that the droplet's initial volume (quantitatively denoted by $\stkout{V}_0$) is covered near-uniformly.

Figure \ref{figure2} plots the cumulative fraction of particles that have intersected the free surface $N_s / N_{\text{tot}}$ as a function of $\theta_c / \theta_{c_0}$ for droplets with initial contact angles $\theta_{c_0} = 30^{\circ}, 60^{\circ}, 90^{\circ}, 120^{\circ}, 150^{\circ}$. We see that, sooner or later, all particles meet the interface (see also the Supplemental Material \cite{supp}). Remarkably, we find that the fraction of intersected particles is almost equal to the fraction of the droplet's lost volume $\left( \stkout{V}_0 - \stkout{V} \right) / \stkout{V}_0$ at a given $\theta_c$ (see Fig.~\ref{figure2}). These observations strongly suggest that any model for the particle deposition should explicitly account for the interaction of solutes with the free surface.

What would happen to the particles after the intersection? There is no experimental evidence supporting that nonvolatile particles cross the interface and enter the gas phase. Thus, they either stay at the interface or are reflected back to the bulk. The latter is also unlikely since, neglecting the particle diffusion, the relative fluid velocity normal to the interface keeps the particles attached. Hence, it is reasonable to assume that the particles are captured by the free surface upon their arrival and are transported along the interface thereafter. Following this assumption, the transport of solutes is divided into two parts: (i) transport in the bulk and (ii) interfacial transport. It is worth noting that the interaction between finite-size particles and fluidic interfaces can be complicated and may lead to scenarios other than the one just described (see, e.g., Refs.~\cite{kaz2012,yunker2011,mcgorty2010}). For instance, the presence of surface-active agents at the liquid-gas interface or particle charge can repel the particles from the interface or facilitate their adsorption. Such complexities are avoided here by only dealing with uncharged pointlike particles and clean interfaces.

Generally, the spatiotemporal evolution of the bulk concentration of solutes $C$ is governed by the advection-diffusion equation
\begin{subequations}\label{eq:bulk_trans_ad}
\begin{align*}
\frac{\partial C}{\partial \theta_c} \dot{\theta}_c
&+ \frac{1}{h_\alpha h_\theta h_\phi}\left[ \frac{\partial }{\partial \alpha}\left( h_\theta h_\phi V_\alpha C \right) + \frac{\partial }{\partial \theta}\left( h_\alpha h_\phi V_\theta C \right) \right]\\
&= \frac{\mathscr{D}}{h_\alpha h_\theta h_\phi}\left[ \frac{\partial }{\partial \alpha}\left( \frac{h_\theta h_\phi}{h_\alpha} \frac{\partial C}{\partial \alpha} \right) + \frac{\partial }{\partial \theta}\left( \frac{h_\alpha h_\phi}{h_\theta} \frac{\partial C}{\partial \theta} \right) \right], \numberthis
\end{align*}
\end{subequations}
which reduces to
\begin{equation}\label{eq:bulk_trans_a}
\frac{\partial C}{\partial \theta_c} \dot{\theta}_c
+ \frac{1}{h_\alpha h_\theta h_\phi}\left[ \frac{\partial }{\partial \alpha}\left( h_\theta h_\phi V_\alpha C \right) + \frac{\partial }{\partial \theta}\left( h_\alpha h_\phi V_\theta C \right) \right] = 0
\end{equation}
when the contribution of diffusion is negligible. Likewise, the transport equation for the surface concentration of captured particles $C_s$ can be written in the toroidal coordinate ($\alpha$,$\theta_c$) as
\begin{subequations}\label{eq:surf_trans_ad}
\begin{align*}
\left[ \frac{\partial}{\partial \theta_c} \left( h_\alpha h_\phi C_s \right) \right] \dot{\theta}_c
&+ \frac{\partial }{\partial \alpha}\left( h_\phi V_\alpha C_s \right)\\
&= h_\alpha h_\phi \dot{C}_s
+ \mathscr{D}_s \frac{\partial }{\partial \alpha}\left( \frac{h_\phi}{h_\alpha} \frac{\partial C_s}{\partial \alpha} \right), \numberthis
\end{align*}
\end{subequations}
where $\dot{C}_s$ is the rate of particle adsorption and the subscript $s$ denotes the surface quantities. A mass balance at the interface indicates that
\begin{equation}\label{eq:adsorption}
\dot{C}_s = \left( V_\theta - V_{\theta_b} \right) C = \left( \frac{J}{\rho} \right) C,
\end{equation}
with $V_{\theta_b}$ and $J$ being the speed at which the boundary is moving in the direction normal to itself and the evaporative flux, respectively (see also Ref. \cite{Masoud2009-a} and the Appendix). Substituting Eq.~(\ref{eq:adsorption}) into Eq.~(\ref{eq:surf_trans_ad}) and neglecting the diffusion term, we obtain
\begin{equation}\label{eq:surf_trans_a_toroid}
\left[ \frac{\partial}{\partial \theta_c} \left( h_\alpha h_\phi C_s \right) \right] \dot{\theta}_c
+ \frac{\partial }{\partial \alpha}\left( h_\phi V_\alpha C_s \right)\\
= h_\alpha h_\phi  \left( \frac{J}{\rho} \right) C,
\end{equation}
where $h_\alpha$, $h_\phi$, $V_\alpha$, and $C$ are all evaluated at $\theta = \theta_c$.

According to Eq.~(\ref{eq:bulk_trans_a}), if we start with $C\left( \alpha, \theta, \theta_{c_0} \right) = \text{const}$, then $\partial C / \partial \theta_c = 0$ due to the incompressibility condition,
\begin{equation}\label{eq:incompressibility}
\nabla \cdot \boldsymbol{V} = \frac{1}{h_\alpha h_\theta h_\phi}\left[ \frac{\partial }{\partial \alpha}\left( h_\theta h_\phi V_\alpha \right) + \frac{\partial }{\partial \theta}\left( h_\alpha h_\phi V_\theta \right) \right] = 0.
\end{equation}
Thus, if the solutes are initially distributed uniformly inside the drop, their bulk concentration remains unchanged during the evaporation. This is in agreement with the results of our particle tracing calculations where the trajectories of uniformly seeded solutes were monitored. We note that neither the no-flux boundary condition at the axis of symmetry and substrate nor the capture of particles by the interface alters the uniformity of the initial distribution. That $C$ remains unaffected as the droplet loses volume might seem counterintuitive. However, one should consider that the number of particles suspended in the bulk also decreases. In fact, as shown in Fig.~\ref{figure2}, the number of particles leaving the bulk (i.e.,~being captured by the interface) is directly proportional to the volume lost by the drop. This can be proven mathematically by integrating both sides of Eq.~(\ref{eq:adsorption}) over the surface of the drop denoted by $S$, i.e.,
\begin{equation}
\int_S \dot{C}_s \, d S
= C \int_0^{2 \pi} \int_0^{\infty} \left( \frac{J}{\rho} \right) h_\alpha h_\phi \, d \alpha \, d \phi
= - C \, \frac{d \stkout{V}}{d t} .
\end{equation}%

Normally, Eqs.~(\ref{eq:bulk_trans_a}) and (\ref{eq:surf_trans_a_toroid}) ought to be solved simultaneously over the lifetime of the drop in order to determine the final distribution of solutes (i.e.,~the deposition pattern). However, based on the foregoing discussion, there is no need to deal with Eq.~(\ref{eq:bulk_trans_a}) (as its solution is already known), and we need to just focus on the interfacial transport of solutes. To this end, we numerically solve Eq.~(\ref{eq:surf_trans_a_toroid}) using an implicit first-order upwind scheme as
\begin{subequations}\label{eq:surf_trans_a_toroid_disc}
\begin{align*}
& \left[ \left( h_\alpha h_\phi C_s \right)_j^{n+1} - \left( h_\alpha h_\phi C_s \right)_j^{n} \right]  \dot{\theta}_c^{n+1} \Delta \alpha + \Delta \theta_c\\
& \times \left\{ \max \left( V_{\alpha_j} / \left|V_{\alpha_j}\right| , 0 \right)
\left[ \left( h_\phi V_\alpha C_s \right)_j^{n+1} - \left( h_\phi V_\alpha C_s \right)_{j-1}^{n+1} \right] \right.\\
& + \left. \min \left( V_{\alpha_j} / \left|V_{\alpha_j}\right| , 0 \right) \left[ \left( h_\phi V_\alpha C_s \right)_{j+1}^{n+1} - \left( h_\phi V_\alpha C_s \right)_j^{n+1} \right] \right\}\\
& = \Delta \theta_c \Delta \alpha \left[ h_\alpha h_\phi  \left( \frac{J}{\rho} \right) C \right]_j^{n+1}, \numberthis
\end{align*}
\end{subequations}
where $\Delta \alpha = \alpha_{j+1} - \alpha_j$ with $j$ being an integer. Again, we use the solution of Masoud and Felske \cite{Masoud2009-b} to evaluate $V_\alpha$ (see also the Appendix). We apply the no-flux condition at $\alpha = 0$ and fix $\Delta \theta_c = -1/2^\circ$, $\Delta \alpha = 0.001$, and $\alpha_\text{max} = 7$. Note that whereas $\alpha$ spans from $0$ to $\infty$, $r/R > 0.998$ for a point with $\alpha > 7$ [see Eq.~(\ref{coordinate_relations})]. It is also assumed that the flux of particles leaving $\alpha = 7$ and the particles adsorbed in the range of $7 < \alpha < \infty$ permanently accumulate at the edge of the drop.

We complement our Eulerian approach with the Lagrangian simulations explained earlier. This time, once a particle intersects the interface we set its $\theta$ coordinate to $\theta_c$ and only track its $\alpha$ coordinate via Eq.~(\ref{disp_toroid_disc}a). For the sake of consistency, we freeze the position of particles whose $\alpha > 7$ and regard them permanently deposited on the substrate (see also the Supplemental Material \cite{supp}). Both types of simulations are carried out down to $\theta_c = 3^{\circ}$ when the contact line is assumed to unpin \cite{Hu2002}. The change in the distribution of particles after the unpinning until the complete drying of the droplet is considered to be insignificant.

\section{Results and discussion}

\begin{figure*}
 \includegraphics[width = \textwidth]{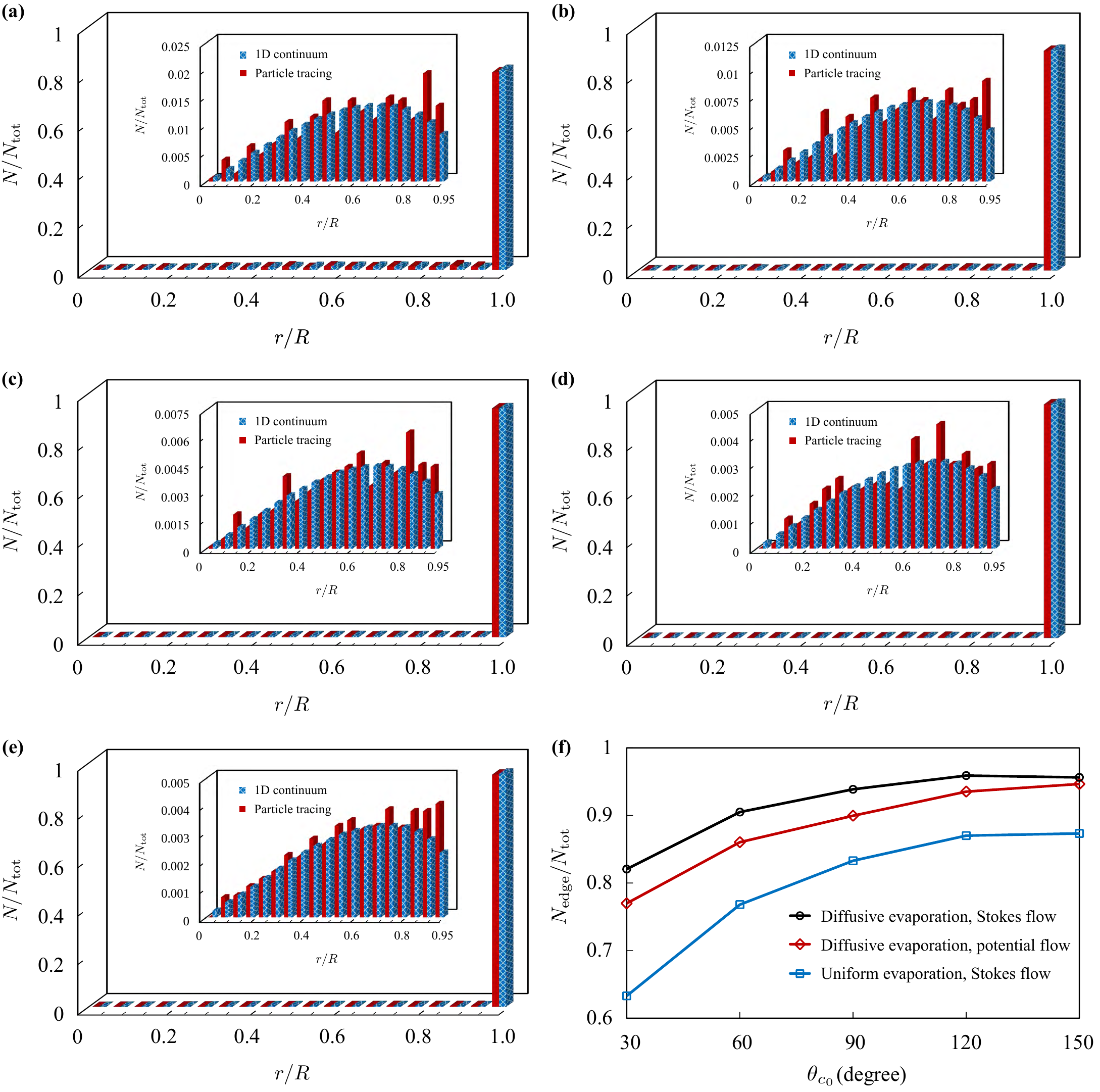}
 \caption{(Colour online) The final distribution of particles in the form of bar charts for droplets with initial contact angles (a) $\theta_{c_0} = 30^\circ$, (b) $\theta_{c_0} = 60^\circ$, (c) $\theta_{c_0} = 90^\circ$, (d) $\theta_{c_0} = 120^\circ$, and (e) $\theta_{c_0} = 150^\circ$. The distance from the axis of symmetry to the contact line is divided into 20 bins and each bar represents the fraction of particles located in the corresponding radial interval. The blue and red bars illustrate the results of the one-dimensional continuum model and particle tracing simulations, respectively. Also, the insets are magnified replots for $0 \le r/R \le 0.95$.  (f) The fraction of particles in the last interval $0.95 < r/R \le 1$ (denoted by $N_{\text{edge}} / N_{\text{tot}}$) as a function of the initial contact angle $\theta_{c_0}$. The circles recap the results of (a)-(e) for the deposition due to the Stokes flow inside the drop induced by a diffusive evaporation profile. The diamonds and squares show particle accumulations at the edge due to, respectively, potential flow induced by the same diffusive evaporation and Stokes flow generated by a uniform evaporative flux. Due to the similarity, only the results of the continuum calculations are shown.}
\label{figure3}
\end{figure*}

We plot the final distribution of particles in the form of bar charts where each bar represents the fraction of particles located in the given interval [see Figs.~\ref{figure3}(a)-\ref{figure3}(e)]. The blue and red bars illustrate the results of the one-dimensional continuum model and particle tracing simulations, respectively. The insets magnify the plots of $N/N_{\text{tot}}$ versus $r/R$ over the range of $0 \le r/R \le 0.95$, where $N$ denotes the count of particles in each bin.

Indeed, we see that the results of the Eulerian and Lagrangian approaches agree very well with each other. More importantly, we find that all patterns predicted by our model are ringlike, i.e., show a large accumulation of particles in the last interval, where $0.95 < r/R \le 1$. We also see that the qualitative form of particle distribution curves for different initial contact angles is very similar. $N/N_{\text{tot}}$ initially increases with increasing $r/R$ until it reaches a maximum at $r/R \sim 0.7$ and then decreases a little before it sharply jumps to a much higher value at the edge. Quantitatively, however, the fraction of particles deposited at the edge $N_{\text{edge}}/N_{\text{tot}}$ is higher for drops with greater  initial contact angles [see also Fig.~\ref{figure3}(f)]. This behavior follows from noting that the tangential component of the surface flow is towards the contact line. Therefore, since drops having larger initial contact angles provide more time for particle motion, a larger fraction of the particles finally deposits at the edge.

\begin{figure}
 \includegraphics[width = 0.445 \textwidth]{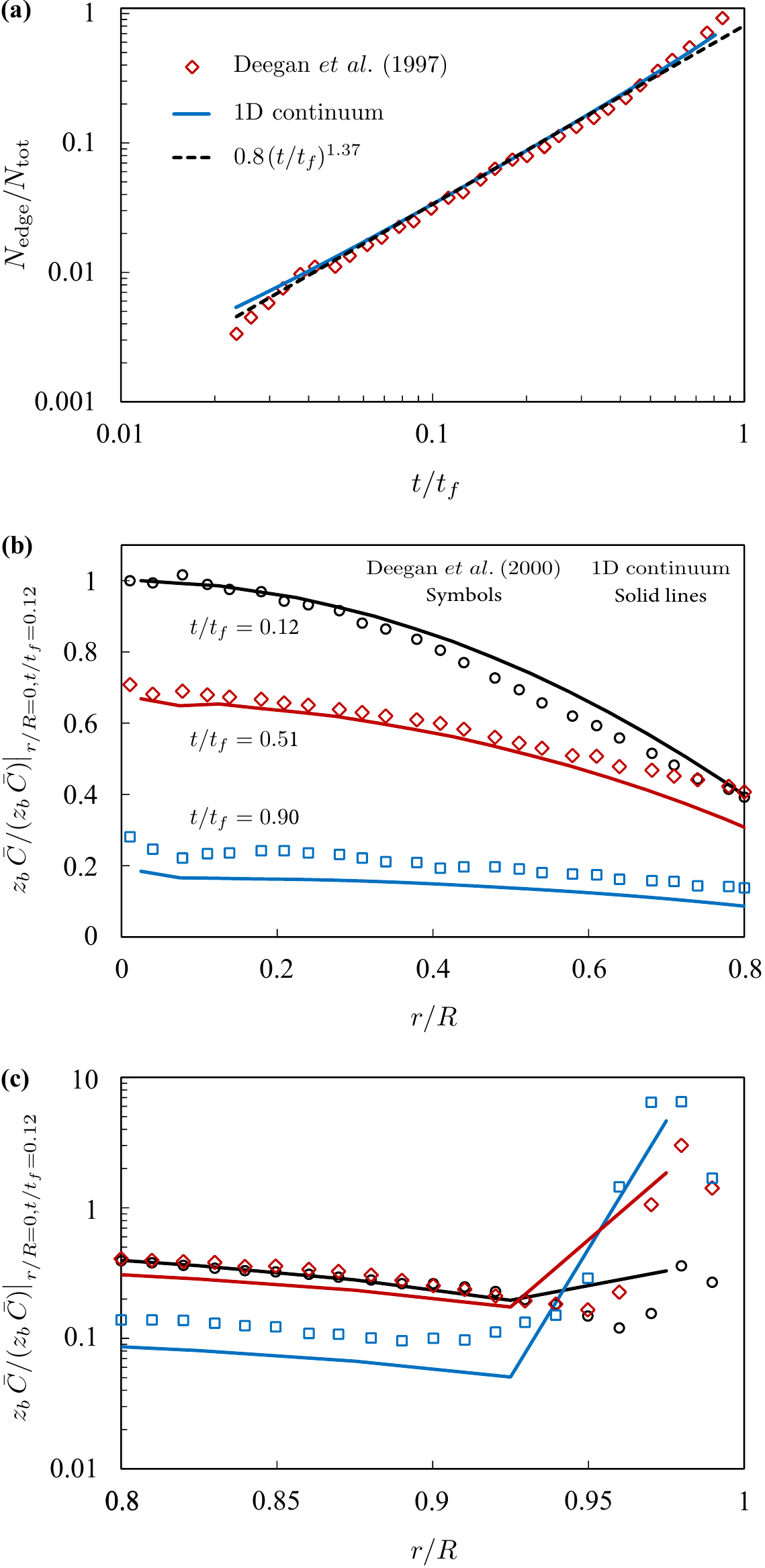}
 \caption{(Colour online) Particle deposition during evaporation of a colloidal droplet with $\theta_{c_0} = 15^\circ$. (a) The cumulative fraction of particles arriving at the contact line $N_{\text{edge}} / N_{\text{tot}}$ versus time $t / t_f$, where $t_f$ is the total drying time whose value for the experiment of Deegan \emph{et al.}~\cite{Deegan1997} is $1300~\unit{s}$. Also, in their experiment the maximum number of particles deposited at the contact line $N_{\text{edge}_\text{max}} = 0.9 N_{\text{tot}}$. To be consistent with the experiments, in this plot, $N_{\text{edge}}$ represents the number of particles in the interval of $0.99 < r / R \le 1$. (b) and (c) Radial distribution of height-averaged concentration of particles at $t / t_f = 0.12,0.51,0.9$. Here, $\bar{C}$ denotes the total (bulk and surface) concentration of solutes averaged over the height of the drop $z_b$ at each radial position, and $z_b \, \bar{C}$ is normalized by its corresponding value at $r / R = 0$ and $t / t_f = 0.12$. Symbols represent the measurements of Deegan \emph{et al.}~\cite{Deegan2000-a} whereas the solid lines denote the predictions of the continuum model. The simulations are carried out down to $\theta_c = 1.5^\circ$ to obtain the results for $t / t_f = 0.9$.}
\label{figure4}
\end{figure}

To further test the validity of our theory, we compare the predictions of the continuum model with the experimental measurements of Deegan \emph{et al.}~\cite{Deegan1997,Deegan2000-a}, who studied particle deposition during the evaporation of colloidal sessile drops with small initial contact angles. In particular, they considered a drop with base radius $R = 2\, \unit{mm}$ and initial contact angle of $\theta_{c_0} = 15^\circ$ containing a dilute suspension of submicron particles, of which a tiny fraction was fluorescent. The fluorescent particles were tracked to count the cumulative number of particles that have deposited at the contact line as a function of time [see Fig.~\ref{figure4}(a)]. In addition, fluorescent video microscopy was used to measure the concentration of solutes throughout the drying time [see Figs.~\ref{figure4}(b) and \ref{figure4}(c)].

Figure \ref{figure4}(a) shows that the results of our model follow the power-law behavior observed experimentally \cite{Deegan1997}. Furthermore, Figs.~\ref{figure4}(b) and \ref{figure4}(c) illustrate that our theory reproduces the overall trend of measurements for the height-averaged distribution of particles \cite{Deegan2000-a}. In these plots, $\bar{C}$ denotes the total (bulk and surface) concentration of solutes averaged over the height of the drop $z_b$ at each radial position, and $z_b \, \bar{C}$ is normalized by its corresponding value at $r / R = 0$ and $t / t_f = 0.12$, where $t_f$ is the total drying time. Notable discrepancies between our results and those of Deegan \emph{et al.}~\cite{Deegan1997,Deegan2000-a} include differences in the fraction of deposited particles at very early times [see Fig.~\ref{figure4}(a)] and the concentration of solutes around $r/R = 0.8$ at $t / t_f = 0.5$ [see Fig.~\ref{figure4}(b)]. Also, the theory underestimates the particle distribution in the range of $0 \le r/R \le 0.9$ at $t / t_f = 0.9$. The kinks in Fig.~\ref{figure4}(c) correspond to the particle accumulation in the last computational bin $0.95 < r/R \le 1$ as seen before in Figs.~\ref{figure3}(a)-\ref{figure3}(e).

The general consistency between our predictions and available experimental measurements \cite{Deegan1997,Deegan2000-a} suggests that the interfacial transport of particles may potentially be an important aspect for the formation of ring-shaped deposition patterns. Needless to say, a more rigorous (and much needed) test of our theory is only possible through comparison with a comprehensive set of experimental data covering a wide range of initial contact angles. We hope that this paper motivates future experiments to collect such a data set.

Encouraged by the agreement shown in Fig.~\ref{figure4} between the results of our model and the previously reported observations, we use the model to predict the effect of the evaporation profile on particle deposition. Figure \ref{figure3}(f) compares the variation of $N_{\text{edge}}/N_{\text{tot}}$ as a function of the initial contact angle $\theta_{c_0}$ for uniform and diffusive evaporative fluxes where the total evaporation rates from the two profiles are identical. Due to the similarity, only the results of the continuum calculations are shown. The comparison indicates that a uniform evaporation profile leads to a less intense accumulation of particles at the edge (i.e., a more uniform deposition pattern). Hence, it is possible to tailor the deposition pattern by imposing a particular distribution of evaporative flux.

A close inspection of previous studies on evaporation-induced flow inside sessile drops \cite{Tarasevich2005,Hu2005-a,Petsi2006,Petsi2008,Masoud2009-a,Masoud2009-b} reveals that Stokes and potential (inviscid) flows are similar near the free surface but are somewhat different close to the substrate due to the no-slip condition in viscous flows. Motivated by this observation and that our model only involves the surface flow, we compare the curves of $N_{\text{edge}}/N_{\text{tot}}$ versus $\theta_{c_0}$ for Stokes and potential flows induced by the same diffusive evaporation profile [see Fig.~\ref{figure3}(f)]. The inviscid surface velocities are calculated based on the analytical solution of Masoud and Felske for potential flow inside evaporating sessile drops \cite{Masoud2009-a}. Because of the differences in the near substrate flow behavior, it might be surmised that Stokes and inviscid flows produce distinctly different deposition patterns. The patterns, however, are actually quite similar [see Fig.~\ref{figure3}(f)]. This is potentially useful since, generally, inviscid velocity fields are easier to calculate (either analytically or numerically) than their viscous counterparts. Therefore, the use of the former in particle deposition studies could reduce the complexity and computational cost of calculations.

\section{Conclusions}

We introduced a theoretical model for the particle deposition during the evaporation of colloidal sessile droplets. The model was founded based on the observation that the suspended particles are captured by the droplet free surface during the course of evaporation. We presented the Eulerian and Lagrangian forms of the model and verified their predictions. The validity of our model indicates that the mechanism of particle capturing by the liquid-gas interface can dominate the transport of particles to the substrate, which highlights the potential role of the free surface in determining the final distribution of particles.

Using the Eulerian description of the model, we showed that the shape of the deposition pattern can be regulated by altering the form of the evaporative flux with more uniform profiles producing more uniform deposition patterns. We also found that the deposition patterns are similar for the viscous and inviscid models of the flow inside the drop. This, of course, results from the fact that, in our model, it is not the bulk flow which controls the deposition but rather the flow at the free surface. It is, therefore, possible to employ more readily calculated inviscid solutions in order to obtain reliable estimates of deposition patterns. Overall, the findings of this paper offer a fresh perspective on the nature of particle deposition during the drying of colloidal drops, which is of high importance for many industrial and scientific processes. We certainly hope that the insight gained by our theoretical analyses guide and/or motivate future experiments on the subject.

Lastly, the theoretical framework presented here can be extended to account for the variation of the shear stress at the free surface due to the Marangoni effect. In fact, the velocity field reported in the Appendix already includes the distribution of shear stress, which was set to zero in our calculations [see Eq.~(\ref{eq:psi_tilde_theta_c})]. Since Marangoni flows have been shown to promote more uniform deposition patterns (see, e.g., Refs.~\cite{sempels2013,still2012,Hu2006}), we expect the extended version of our model to reproduce this behavior as well. Additionally, the framework can readily be used for two-dimensional drops \cite{Yarin2006} by writing the equations in the bipolar coordinates instead of the toroidal (see, e.g., Ref.~\cite{Masoud2009-b}).

\section*{Appendix: Velocity field inside an evaporating sessile drop}

Masoud and Felske analytically solved the axisymmetric Stokes flow within an evaporating sessile drop of spherical cap shape \cite{Masoud2009-b}. Solutions are obtained for arbitrary contact angles and evaporative flux distributions along the free surface. They considered droplets whose contact lines are either pinned or free to move during the evaporation. Following their derivations for pinned contact lines, the components of the evaporation-induced flow in the toroidal coordinates can be written as
\begin{mysubequations}\label{eq:V_alpha}
\begin{align*}
V_\alpha = \frac{ \sqrt{ \cosh \alpha + \cos \theta } } { R^2\sinh \alpha } \left[ \frac{3 \sin \theta} {2} \sqrt{ \cosh \alpha + \cos \theta } \, \psi \left( \alpha ,\theta, \theta_c \right) \right.\\
\left. + \int_{0}^{\infty} \frac { \partial K \left(\theta ,\tau, \theta_c \right) } { \partial \theta } \, C_{1/2 + i \tau }^{-1/2} \left( \cosh \alpha \right) d \tau \right], \numberthis
\end{align*}
\end{mysubequations}%
\begin{mysubequations}\label{eq:V_theta}
\begin{align*}
V_\theta = \frac{ \sqrt{ \cosh \alpha + \cos \theta } } { R^2 } \left[ \frac{3 \sqrt{ \cosh \alpha + \cos \theta }} {2}  \, \psi \left( \alpha ,\theta, \theta_c \right) \right.\\
\left. + \int_{0}^{\infty} K \left(\theta ,\tau, \theta_c \right) \, P_{-1/2 + i \tau } \left( \cosh \alpha \right) d \tau \right], \numberthis
\end{align*}
\end{mysubequations}%
where $C_{1/2 + i\tau }^{-1/2}$ is the Gegenbauer function of the first kind and of order $-1/2$, and $P_{-1/2 + i \tau } \left( x \right)$ is the conical function of the first kind. Here, the stream function $\psi \left( \alpha ,\theta, \theta_c \right)$ is calculated from
\begin{mysubequations}\label{eq:psi}
\begin{align*}
\psi \left( \alpha ,\theta, \theta_c \right) &= \left( \cosh \alpha + \cos \theta \right)^{-3/2}\\
&\times \int_{0}^{\infty} K \left( \theta ,\tau, \theta_c \right) \, C_{1/2 + i\tau }^{-1/2} \left( \cosh \alpha \right) d \tau, \numberthis
\end{align*}
\end{mysubequations}%
where
\begin{mysubequations}\label{eq:K}
\begin{align*}
K(\theta ,\tau, \theta_c ) &= k_1 \left( \tau, \theta_c \right) \sin \theta \sinh \left( \tau \, \theta \right)\\
&+ k_2 \left(\tau, \theta_c \right) \left[ \cos \theta \sinh \left( \tau \, \theta \right) - \tau \sin \theta \cosh \left( \tau \, \theta \right) \right]. \numberthis
\end{align*}
\end{mysubequations}%
The functions $k_1\left( \tau, \theta_c \right)$ and $k_2\left( \tau, \theta_c \right)$ are obtained via
\begin{myequation}\label{eq:k1}
k_1 \left( \tau, \theta_c \right) = \frac { N_2 \left(\tau ,\theta_c \right) K \left(\tau ,\theta_c \right) + N_1 \left( \tau, \theta _c \right) \widetilde{K} \left( \tau , \theta_c \right) } { N_2 \left( \tau ,\theta_c \right) M_1 \left( \tau ,\theta_c \right) + N_1 \left( \tau ,\theta_c \right) M_2 \left( \tau ,\theta_c \right) },
\end{myequation}%
\begin{myequation}\label{eq:k2}
k_2 \left( \tau, \theta_c \right) = \frac { M_2 \left(\tau ,\theta_c \right) K \left(\tau ,\theta_c \right) - M_1 \left( \tau, \theta _c \right) \widetilde{K} \left( \tau , \theta_c \right) } { N_2 \left( \tau ,\theta_c \right) M_1 \left( \tau ,\theta_c \right) + N_1 \left( \tau ,\theta_c \right) M_2 \left( \tau ,\theta_c \right) },
\end{myequation}%
with
\begin{myequation}\label{eq:M1}
M_1 \left( \tau ,\theta_c \right) = \sin \theta_c \sinh \left( \tau \, \theta_c \right),
\end{myequation}%
\begin{mysubequations}\label{eq:M2}
\begin{align*}
M_2 \left( \tau ,\theta_c \right) &= \left( \tau^2 - 1 \right) \sin \theta_c \sinh \left( \tau \, \theta_c \right)\\
&+ 2 \tau \cos \theta_c \cosh \left( \tau \, \theta_c \right), \numberthis
\end{align*}
\end{mysubequations}%
\begin{myequation}\label{eq:N1}
N_1 \left( \tau ,\theta_c \right) = \cos \theta \sinh \left(\tau \, \theta \right) - \tau \sin \theta \cosh \left( \tau \, \theta \right),
\end{myequation}%
\begin{mysubequations}\label{eq:N2}
\begin{align*}
N_2 \left( \tau ,\theta_c \right) = &\left(\tau^2 + 1 \right)\\
&\times \left[ \cos \theta_c \sinh \left( \tau \, \theta_c \right) + \tau \sin \theta_c \cosh \left( \tau \, \theta_c \right) \right]. \numberthis
\end{align*}
\end{mysubequations}%

The functions $K \left(\tau ,\theta_c \right)$ and $\widetilde{K} \left(\tau ,\theta_c \right)$ are related to the boundary conditions at the free surface of the drop as
\begin{mysubequations}\label{eq:K_theta_c}
\begin{align*}
&K \left(\tau ,\theta_c \right) = \tau \left( \tau^2 + \frac{1}{4} \right) \tanh \left( \pi \, \tau \right)\\
&\times \int_0^{\infty} \frac{ \psi \left( \alpha , \theta_c \right) \left( \cosh \alpha + \cos \theta_c \right)^{3/2} } {\sinh \alpha} \, C_{1/2 + i\tau }^{-1/2} \left(\cosh \alpha \right) d \alpha, \numberthis
\end{align*}
\end{mysubequations}%
\begin{mysubequations}\label{eq:K_tilde_theta_c}
\begin{align*}
\widetilde{K} \left(\tau ,\theta_c \right) &= \tau \left( \tau^2 + \frac{1}{4} \right) \tanh \left( \pi \, \tau \right)\\
&\times \int_0^{\infty} \frac{ \widetilde{\psi} \left( \alpha , \theta_c \right)} {\sinh \alpha} \, C_{1/2 + i\tau }^{-1/2} \left(\cosh \alpha \right) d \alpha, \numberthis
\end{align*}
\end{mysubequations}%
where
\begin{mysubequations}\label{eq:psi_theta_c}
\begin{align*}
\psi \left(\alpha ,\theta_c \right) = &- \frac{R^3 \dot{\theta}_c}{2} \left[ \frac{1}{\left( 1 + \cos \theta_c \right)^2} - \frac{1}{\left( \cosh \alpha + \cos \theta_c \right)^2} \right]\\
&- \int_0^{\alpha} \frac{ R^2 \, \sinh \alpha^\prime} {\left( \cosh \alpha^\prime + \cos \theta_c \right)^2} \frac { J \left(\alpha^\prime, \theta_c \right)} {\rho} \, d \alpha^\prime, \numberthis
\end{align*}
\end{mysubequations}%
\begin{mysubequations}\label{eq:psi_tilde_theta_c}
\begin{align*}
&\widetilde{\psi} \left(\alpha ,\theta_c \right) = - \left[ \frac{R^2 \sinh \alpha} {\left( \cosh \alpha + \cos \theta_c \right)^{3/2}} \right]\\
&\times \left\{ \frac{\tau_{\alpha \theta} R } {\mu} + \frac{\partial}{\partial \alpha} \left[ \left( \cosh \alpha + \cos \theta_c  \right) V_\theta \left(\alpha, \theta_c \right) \right] \right\}\\
&- \frac{3 \psi \left(\alpha ,\theta_c \right)}{2} \left[ \frac{\cos \theta_c \left( \cosh \alpha + \cos \theta_c  \right) - \frac{1}{2} \sin^2 \theta_c} {\sqrt{ \cosh \alpha + \cos \theta_c}} \right]. \numberthis
\end{align*}
\end{mysubequations}%
Here, $J \left(\alpha, \theta_c \right)$ is the evaporative flux, $\rho$ is the liquid density, $\tau_{\alpha \theta}$ is the shear stress at the interface (which is zero in the absence of Marangoni effects), and
$\dot{\theta}_c$ is the rate of change of the contact angle. The mass balance at the boundary requires that
\begin{myequation}\label{eq:V_theta_theta_c}
V_\theta \left(\alpha ,\theta_c \right) = \frac{R \dot{\theta}_c}{\cosh \alpha + \cos \theta_c} + \frac { J \left(\alpha, \theta_c \right)} {\rho}.
\end{myequation}%
Also, the rate of volume loss by the drop results in
\begin{mysubequations}\label{eq:dtheta_c_dt}
\begin{align*}
\dot{\theta}_c = \frac{d \theta_c}{dt} = &- \frac{2 \left( 1 + \cos \theta_c \right)^2} {R \rho}\\
&\times \int_0^{\infty} \frac{\sinh \alpha \, J \left(\alpha, \theta_c \right)} {\left( \cosh \alpha + \cos \theta_c \right)^2} \, d \alpha. \numberthis
\end{align*}
\end{mysubequations}%
For a given $J$, the flow within the drop can be obtained by substituting Eqs.~(\ref{eq:psi})-(\ref{eq:dtheta_c_dt}) into Eqs.~(\ref{eq:V_alpha}) and (\ref{eq:V_theta}).

Two commonly considered evaporations correspond to uniform flux [i.e., $J \left(\alpha, \theta_c \right) = \text{const} = J_0$] and to purely diffusive gas phase transport into an infinite ambient. The latter takes the form of
\begin{mysubequations}\label{eq:J_diff}
\begin{align*}
J \left(\alpha ,\theta_c \right) & = \frac{D_v \left(\varphi_s - \varphi_\infty \right)}{R} \left\{ \frac{\sin \theta_c}{2} \right.\\
& + \sqrt{2} \left( \cosh \alpha + \cos \theta_c \right)^{3/2} \int_0^\infty \frac{\cosh \left( \tau \theta_c \right)} {\cosh \left(\pi \tau \right)} \\
&\times \left. \tanh \left[ \left( \pi - \theta_c \right) \tau \right] \, P_{-1/2 + i\tau } \left (\cosh \alpha \right) \, \tau \, d\tau \right\}, \numberthis
\end{align*}
\end{mysubequations}%
where $D_v$ is the vapor diffusion coefficient, $\varphi_s$ is the vapor density at the droplet surface (saturation value), and $\varphi_\infty$ is the far field vapor density \cite{Popov2005,Masoud2009-a}. Note that the singular behavior of $J$ in Eq.~(\ref{eq:J_diff}) (which takes place in the limit $\alpha \rightarrow \infty$ for $0 \le \theta_c < \pi/2$) needs to be remedied in order to obtain uniformly valid solutions for the liquid motion (see Ref. \cite{Masoud2009-b} for more details).

\bibliographystyle{apsrev}
\bibliography{pre}

\end{document}